\documentclass[11pt,preprint]{aastex}
\newcommand{\scm}{cm$^{-2}$}
\newcommand{\cc}{cm$^{-3}$}
\newcommand{\epsC}{$\epsilon$~CMa}

\newcommand{\ACAS}{$A_\mathrm{C}/A_\mathrm{S}$}
\newcommand{\NH}{$N(\mathrm{H})$}
\newcommand{\kms}{km s$^{-1}$}

\newcommand{\NII}{\ion{N}{2}}

\newcommand{\NI}{\ion{N}{1}}

\begin{document}
\title{Evidence for a High Carbon Abundance in the Local Interstellar Cloud}

\author{Jonathan D. Slavin\altaffilmark{1} and Priscilla C.
Frisch\altaffilmark{2}}
\altaffiltext{1}{Harvard-Smithsonian Center for Astrophysics, 60 Garden Street,
MS 83, Cambridge, MA 02138}

\altaffiltext{2}{University of Chicago, Department of Astronomy and
Astrophysics, 5460 S.\ Ellis Avenue, Chicago, IL 60637}

\begin{abstract}

The nature of the Local Interstellar Cloud (LIC) is highly constrained by the
combination of \emph{in situ} heliospheric and line-of-sight data towards
nearby stars. We present a new interpretation of the LIC components of the
absorption line data towards \epsC, based on recent atomic data that include
new rates for the Mg$^+$ to Mg$^0$ dielectronic recombination rate, and using
\emph{in situ} measurements of the temperature and density of neutral helium
inside of the heliosphere.  With these data we are able to place interesting
limits on the gas phase abundance of carbon in the LIC.  If the C/S abundance
ratio is solar, $\sim 20$, then no simultaneous solution exists for the
$N($\ion{Mg}{1}$)$, $N($\ion{Mg}{2}$)$, $N($\ion{C}{2}$)$ and
$N($\ion{C}{2}$^*)$ data. The combined column density and \emph{in situ} data
favor an abundance ratio \ACAS\ $= 47^{+22}_{-26}$.

We find that the most probable gas phase C abundance is in the range 400 to
800 ppm with a lower limit of $\sim 330$.  We speculate that such a supersolar
abundance could have come to be present in the LIC via destruction of
decoupled dust grains.  Similar enhanced C/H ratios are seen in very low
column density material, \NH$ < 10^{19}$ \scm, towards several nearby stars.
\end{abstract}


\section{Introduction} \label{sec:intro}

The interstellar cloud that surrounds our solar system, the 
Local Interstellar Cloud (LIC), is a tiny, radius $\lesssim 1$ pc, low
%density, $n\sim0.3$ \cc, warm, $T \approx 6300$ K, interstellar cloud that has
density interstellar cloud that has
been detected in absorption lines towards many towards nearby stars
\citep[e.g.,][]{McClintocketal:1978,Bruhweiler:1984,Genovaetal:1990,
Lallement+Ferlet_1997,Hebrard_etal_1999,Gry+Jenkins_2001,
Wood+Linsky_1997}.  The cloud is partially ionized, $\sim 20-40$\%, apparently
due to the low intensity, relatively hard ionizing radiation field
\citep[][hereafter SF02]{Slavin+Frisch_2002,FrischSlavin:2003}.  
Direct observations of interstellar He$^0$ which penetrates the heliosphere
gives us tight constraints on its temperature and density: $T = 6300\pm340$ K,
$n($He$^0) = 0.015\pm0.0015$ cm$^{-3}$ \citep{Witte_2004,Moebius_etal_2004}.

The star $\epsilon$ CMa (HD 52089) is located in the downwind direction of the
flow of ISM past the Sun ($\ell = 239.8$\degr, $b = -11.3$\degr, $d = 132$
pc).  The absorption line data set for the line of sight towards \epsC\ is
particularly complete \citep[][hereafter GJ]{Gry+Jenkins_2001}.  Using this
dataset, in particular the ratios $N($\ion{Mg}{2}$)/N($\ion{Mg}{1}$)$ and
$N($\ion{C}{2}$^*)/N($\ion{C}{2}$)$, GJ put constraints on the electron
density, $n(e)$, and temperature of the LIC and of the other clouds seen in
this line of sight.  As we discuss, their method involved assuming a gas phase
abundance ratio for C/S of 20, appropriate to Solar abundances.  However, we
show that when updated atomic rates are used for the Mg$^+$ recombination
coefficient, in particular for dielectronic recombination, this solution
becomes no longer viable.  This leads us to the conclusion that C has a
supersolar abundance in the LIC, $A_\mathrm{C} =  400 - 800$ ppm (parts per
million).  This result, which is consistent with detailed models of the
photoionization of the LIC \citep[][hereafter SF06]{Slavin+Frisch_2006}, then
allows for a solution for $n(e)$ which is consistent with all other data.
High carbon abundances are also seen towards several other stars that sample
the LIC.

\section{Data}

GJ presented data obtained with the GHRS on HST, and IMAPS on ORFEUS, of the
absorption lines from a number of ions observed towards \epsC, including
\ion{Mg}{1}, \ion{Mg}{2}, \ion{C}{2}, \ion{C}{2}$^*$ and \ion{S}{2}.  Multiple
velocity components are seen in this line of sight, but here we concentrate on
the LIC component, referred to as component 1 by GJ, at a heliocentric
velocity of $\sim 17$ \kms.  The derived column densities, along with
the errors estimated by GJ, are listed in Table \ref{tab:coldens}.  GJ
determined a lower limit on $N($\ion{C}{2}$)$ directly from the observed
absorption line, but needed to rely on an assumption for the abundance ratio
of C to S to derive the upper limit on the column density because of the
degree of saturation of the 1335 \AA\ \ion{C}{2} line.  From these data, using
pre-1982 atomic rate constants, GJ obtained $T=5700-8200$ K and
$n(e)=0.08-0.17$ \cc.

For the LIC we have the additional constraint given by \emph{in situ}
observations of the temperature of inflowing He$^0$ determined from direct
measurements inside of the solar system by the GAS instrument on the Ulysses
space craft, $T = 6300\pm340$ K \citet[][see also
\citealt{Moebius_etal_2004}]{Witte_2004}.  Because of the small column density
of the LIC along the \epsC\ line of sight, $N(\mathrm{H\;I}) \approx (3 -
5)\times 10^{17}$ \scm, the gas temperature is not expected to vary greatly
along the line of sight.  In our photoionization models in which we use a
realistic ionizing radiation field (SF02, SF06), the typical total variation
in temperature is $\sim 10$\%.  We note that if GJ had required consistency
with the LIC temperature in their analysis, their minimum electron density
would have increased to $n(e) \sim 0.11$ \cc\ (up from 0.08 \cc).  

\section{Mg Ionization}
In ionization equilibrium the density ratio of Mg$^+$/Mg$^0$ is determined by
\begin{equation}
\left[\Gamma(\mathrm{Mg}^0) + C^\mathrm{CT}(\mathrm{Mg}^0)
n(\mathrm{H}^+)\right] n(\mathrm{Mg}^0) = \alpha(\mathrm{Mg}^+)\, n(e)\,
n(\mathrm{Mg}^+),
\label{eq:mgionbal}
\end{equation}
where $\Gamma(\mathrm{Mg}^0)$ is the photoionization rate,
$C^\mathrm{CT}(\mathrm{Mg}^0)$ is the charge transfer ionization rate for
Mg$^0$ with H$^+$, $\alpha(\mathrm{Mg}^+)$ is the total Mg$^+$ recombination
coefficient (including both radiative and dielectronic recombination), and
$n(e)$ is the electron density.  The dominant contribution to the
photoionization rate for Mg$^0$ is the FUV background, which comes mostly from
A and B stars.  The spectrum and flux of the background has been estimated by
a number of authors.  We choose to use the results of
\citet{Gondhalekar_etal_1980} because they make use of the TD-1 observations,
which remain the best direct measurements of the background from stars and
includes the whole sky.  This is important because the sky is extremely
anisotropic in the FUV and other observations have covered only brighter than
average portions of the sky.  Integrating the \citet{Gondhalekar_etal_1980}
spectrum over the Mg$^0$ photoionization cross section leads to a value of
$\Gamma(\mathrm{Mg}^0) = 4.5\times10^{-11}$ s$^{-1}$.  The uncertainty of the
FUV spectrum is estimated by \citet{Gondhalekar_etal_1980} to be 25\%.  The
Mg$^0$ photoionization cross section we use comes from
\citet{Verner_etal_1996} and is based on results from the Opacity Project.
Our value for the photoionization rate is significantly below the value $6.1
\times 10^{-11}$ s$^{-1}$ used by GJ, which was based on estimates by
\citet{Jura_1974} and \citet{Mathis_etal_1983}.  The rate of charge transfer
ionization of Mg$^0$ is taken from the fit provided in
\citet{Kingdon+Ferland_1996}, which was based on data from
\citet{Allan_etal_1988}, and gives $C^\mathrm{CT}(\mathrm{Mg}^0)=
6.5\times10^{-11}$ cm$^3\,$s$^{-1}$ at $T = 6300$ K.

The recombination rates, both dielectronic and radiative, are taken from new
calculations.  The radiative rates we use are recently calculated by
Badnell\footnote{\url{http://amdpp.phys.strath.ac.uk/tamoc/DATA/RR/}} using
the AUTO\-STRUCTURE code.  These rates are considerably larger than those
calculated by \citet{Aldrovandi+Pequignot_1973}. We use the dielectronic
recombination rates calculated by this group as well \citep{Altun_etal_2006},
which are also larger than the older combined low temperature
\citep{Nussbaumer+Storey_1986} and high temperature \citep{Burgess_1965}
rates.  The resulting total Mg$^+$ recombination rate for the LIC temperature
range is $\alpha(\mathrm{Mg}^+) = 1.71^{+0.43}_{-0.34}\times 10^{-12}$ cm$^3$
s$^{-1}$.  As a result of the higher rates, the electron density inferred for
a given observed ratio $N($\ion{Mg}{2}$)/N($\ion{Mg}{1}$)$ is considerably
lower than if the older rates had been used.  It should be emphasized that
there are considerable uncertainties in these theoretical dielectronic
recombination rates. The low temperature ($<10^4$ K) dielectronic
recombination rates in particular need to be confirmed by laboratory
measurements and further theoretical calculations.  Since we do not have any
way of estimating the magnitude of the uncertainties in the rates we do not
include the uncertainties in our numerical results and plots.  Some
comfort may be taken from the fact that other recently calculated rates
\citep{Gu_2004} are within $\sim 30$\% of the \citet{Altun_etal_2006} results
for temperatures of interest, though this does not guarantee their accuracy to
that level.

To use equation (\ref{eq:mgionbal}) to solve for the electron density, one
needs to know $n(\mathrm{H}^+)$.  Based on our ionization models (SF02) and the
observation that $N($\ion{H}{1}$)/N($\ion{He}{1}$) \gtrsim 12$
\citep{Dupuis_etal_1995} in the local ISM, we assume that $n(\mathrm{H}^+) =
0.9 n(e)$. This should be good to about 10\% and in any case the charge
transfer term is small compared with the recombination term in the equation
for the electron density.

\section{C$^+$ Fine Structure Excitation}

The ratio of C$^+$ ions in the excited $J=3/2$ fine structure level of the
ground state to that in the $J=1/2$ level is another measure of $n(e)$. 
The equilibrium populations of the two states are given by
\begin{equation}
\frac{n(\mathrm{C}^{+*})}{n(\mathrm{C}^{+})} = \frac{\gamma_{12} n(e) +
\gamma_{12}^H n(\mathrm{H}^0)} {A_{21} + \gamma_{21} n(e) + \gamma_{21}^H
n(\mathrm{H}^0)},
\label{eq:cionbal}
\end{equation}
where $A_{21} = 2.29\times10^{-6}$ s$^{-1}$ is the radiative decay
probability, and $\gamma_{12}$ and $\gamma_{21}$ are the 
excitation and de-excitation coefficients respectively for collisions with
electrons and $\gamma_{12}^H$ and $\gamma_{21}^H$ are the corresponding
coefficients for excitation by neutral H.  The electron collisional excitation
coefficient is 
$\gamma_{12} = \frac{8.63\times10^{-6}}{g_1 T^{1/2}} \Omega_{12}
\exp\left(-\frac{E_{12}}{k T}\right)$,
where $g_1$ is the statistical weight ($=2$) of the lower level, $\Omega_{12}$
is the collision strength and $E_{12}$ is the excitation energy \citep[see,
e.g.][]{Spitzer_1978}.  For partially ionized warm gas such as in the LIC the
excitation of C$^+$ by neutral H is small compared with excitation by
electrons but for completeness we include it using rates from
\citet{Keenan_etal_1986} and assuming a H$^0$ density of 0.2 \cc, which should
be roughly correct.  We use the collision strength from
\citet{Blum+Pradhan_1992}, which is substantially smaller ($\Omega_{12}
\approx 2.0$ at $T = 6300$ K) than the \citet{Hayes+Nussbaumer_1984} value
used by GJ (2.81).  This smaller value is consistent within 5\% with recent
calculations by \citet{Wilson+Bell_2002}.  A smaller value for $\Omega_{12}$
leads to a larger value for $n(e)$ estimated from
$N($\ion{C}{2}$^*)/N($\ion{C}{2}$)$.  For $T=6300$ K, $\gamma_{12}
=1.07\times10^{-7}$ cm$^{3}\,$s$^{-1}$.

\section{Results}

\begin{figure}
\plotone{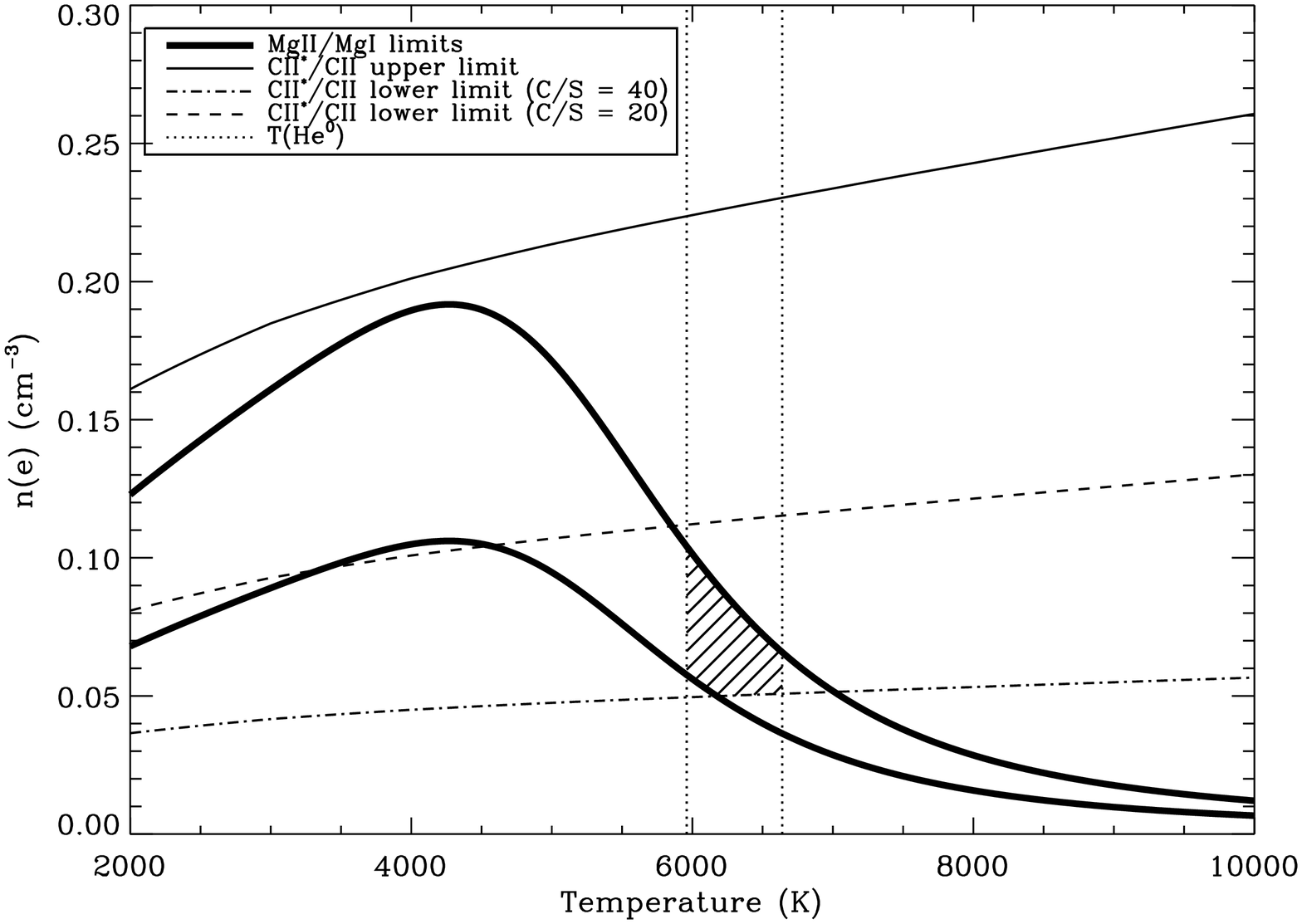}
\caption{Constraints on the electron density and temperature in the LIC based
on column density measurements towards \epsC\ and \emph{in situ} observations
of the temperature. The thick lines show the constraints from the observations
of \ion{Mg}{1} and \ion{Mg}{2}.  The upper, thin solid line is the upper limit
on the electron density using the lower limit on $N($\ion{C}{2}$)$ derived
directly from the observations.  The dashed and dashed-dot lines show the
lower limit on $n(e)$ that result from an upper limit on $N($\ion{C}{2}$)$
under the assumptions of a C/S abundance ratio of 20 and 40 respectively. The
vertical dotted lines indicate the temperature limits based on \emph{in situ}
observations of He$^0$.\label{fig:GJ_plot}}
\end{figure}

Following GJ, we determine $n(e)$ from the overlap region that is consistent
with both $N($\ion{C}{2}$^*)/N($\ion{C}{2}$)$ and
$N($\ion{Mg}{1}$)/N($\ion{Mg}{2}$)$.  The updated atomic data we use leads to
significantly different values from those found by GJ. Figure
\ref{fig:GJ_plot} illustrates the allowed region for $n(e)$ as a function of
$T$, based on equations (\ref{eq:cionbal}) and (\ref{eq:mgionbal}) and the LIC
temperature constraints.  The allowed range of $n(e)$ is much smaller than
found previously primarily because of the limits on the temperature provided
by the \emph{in situ} observations.  As can be seen from equation
(\ref{eq:cionbal}), the lower limit on $n(e)$ from
$N($\ion{C}{2}$^*)/N($\ion{C}{2}$)$ comes from the lower limit on the observed
ratio.  In the figure the dashed line, labeled ``\ion{C}{2}$^*$/\ion{C}{2}
(C/S = 20)'', uses an upper limit on $N($\ion{C}{2}$)$ derived from taking the
upper limit on $N($\ion{S}{2}$)$ and an assumed C/S abundance ratio of 20.
Clearly this value does not allow for a solution consistent with the Mg ion
column densities and LIC temperature.  The lower curve corresponds to a
$N($\ion{C}{2}$)$ upper limit derived from assuming a C/S abundance ratio of
40, and allows a solution with the resulting limits of $n(e) = 0.050 - 0.104$
\cc\ when T is limited by the \emph{in situ} He data.  While C/S need not be
as high as 40 to find an allowed region in the plot, a value of $\gtrsim 21$
is required.

An alternative method of placing constraints on the C/S abundance ratio is to
use $N($\ion{Mg}{1}$)$, $N($\ion{Mg}{2}$)$ and $T$ to derive $n(e)$, and then
$N($\ion{C}{2}$^*)$ and $N($\ion{S}{2}$)$ to find \ACAS\ ($A_\mathrm{C} \equiv
\mathrm{C/H}$ and $A_\mathrm{S} \equiv \mathrm{S/H}$).  Combining equations
(\ref{eq:mgionbal}) and (\ref{eq:cionbal}), we find
\begin{equation}
\frac{A_\mathrm{C}}{A_\mathrm{S}} = \frac{A_{21} + \gamma_{21}
n(e)}{\gamma_{12} n(e)} \frac{N(\mathrm{C\;II}^*)}{N(\mathrm{S\;II})}
\label{eq:ACAS}
\end{equation}
where $n(e)$ is derived via equation (\ref{eq:mgionbal}).  
Observations with reasonable uncertainties exist for all of the quantities
needed to determine \ACAS.  A straightforward propagation of errors
approach yields \ACAS\ $= 46.8\pm25.7$ (including uncertainties in the
temperature and its effect on the rate coefficients). The uncertainties
involved in the expressions contribute in nonlinear ways, however, so we have
in addition carried out a Monte Carlo simulation to estimate the errors.  To
do this we have used the 1-$\sigma$ reported errors for the observed
quantities, assumed normally distributed errors and calculated the resulting
\ACAS\ from equation \ref{eq:ACAS} for many trials.  The resulting probability
distribution is very asymmetric with a long tail to high values of \ACAS.  The
result is \ACAS\ $= 46.8^{+22.0}_{-26.2}$, where the error limits are those
that contain 68.3\% of the probability surrounding the peak in the
distribution.  The peak of the probability distribution function is at 37.2
and the mean (expected value) is 58.5.  Alternatively we may state that the
\ACAS\ $> 21$ with 95.45\% confidence or \ACAS\ $> 39$ with 68.3\% confidence
(note that the long tail to high values causes this one-sided limit to exceed
the value at the peak of the probability distribution). The high-value tail is
a standard result for the ratio of two values in which the errors for each
quantity are a relatively large fraction of the quantity.  The
inclusion of the uncertainty in the Mg$^0$ photoionization rate broadens the
uncertainty range in $n(e)$ beyond that indicated in Figure \ref{fig:GJ_plot}.
The Mg ionization calculations also yield the probability distribution for
electron density for which we find $n(e) = 0.064^{+0.021}_{-0.036}$, where
again we report the limits that enclose 68.3\% of the probability.  For $n(e)$
the peak is at 0.052 \cc\ and the expected value is 0.067 \cc.

Determinations of the electron density for the LIC using independent sets of
data, combined with models, lead \citet{Izmodenov_etal_2003}, to find $n(e) =
0.07\pm0.02$.  Similar calculations by \citet{GloecklerGeiss:2004}, but with
somewhat different assumptions gave $n(e) = 0.049\pm0.016$.  

Another approach for evaluating the C abundance, that is separate from the S
abundance, uses only the more reliable $N($\ion{C}{2}$^*)$ data.  GJ placed
limits on $N($\ion{H}{1}$)$ and $N(\mathrm{H}_\mathrm{tot})$ using several
assumptions.  Using the fact that the ionization of O and H are tightly
coupled by charge transfer \citep{Field+Steigman_1971}, observations of
\ion{O}{1} and an assumed gas phase abundance of 316 ppm, they find
$N($\ion{H}{1}$) = 4.4^{+1.6}_{-0.6}\times10^{17}$ \cc. A recent determination
of O/H in the Local Bubble \citep[][also see
\citet{Andreetal:2003}]{Oliveira_etal_2005} gives a value of $345\pm19$ ppm,
which leads to a somewhat smaller total \ion{H}{1} column density,
$N($\ion{H}{1}$) = 4.0^{+1.5}_{-0.6}\times10^{17}$ \cc.  

The ionization fraction of H in the LIC can be estimated in various ways.
Using just the directly measured EUVE flux from stars, a lower limit of
$X(\mathrm{H}^+) \ge 0.15$ can be determined.  This flux is inadequate to
explain the high He ionization fraction relative to H found from nearby white
dwarf stars, however.\footnote{For instance, the nearby white dwarf stars
GD\,71 and HZ\,43 have $\chi(\mathrm{He})=$ \ion{He}{2}/(\ion{He}{1} +
\ion{He}{2}) $=0.31$, and 0.38 respectively \citep[][and references
therein]{Wolffetal:1999}.} Our models for the ionization of the LIC, which
include the diffuse soft X-ray background as an ionization source and a
contribution from radiation generated in an evaporative boundary between the
LIC and the hot gas of the Local Bubble, predict
ionization fractions of $22 - 36$\%.  For our best new models 
we find total H columns ranging from $4.1 - 5.5\times10^{17}$. While
we cannot place firm upper limits on the H ion fraction, it appears unlikely
that it significantly exceeds 40\% unless some additional unknown ionizing
source is present or the cloud is out of photoionization equilibrium (see
discussion below).  Combining our H ionization fraction limits with the
\ion{H}{1} column determination leads to $N(\mathrm{H}_\mathrm{tot}) \approx
5.6^{+3.0}_{-1.2}\times10^{17}$ \scm.  This value yields a value for the S
abundance of $A_\mathrm{S} \sim 15$ ppm, with 50\% uncertainties, in agreement
with solar values of $12.3 - 15.5$ \citep[][AGS]{Asplund_etal_2005}.  

We can express the C abundance as a function of the observed value of
$N($\ion{C}{2}$^*)$ as,
\begin{equation} 
A_\mathrm{C} = \frac{N(\mathrm{C\,II}^*)}{N(\mathrm{H}_\mathrm{tot})}
\frac{A_{21}}{\gamma_{12} n(e)} =
\left(\frac{N(\mathrm{C\,II}^*)}{1.3\times10^{12}\,\mathrm{cm}^{-2}}\right)
\left(\frac{5\times10^{17}\,\mathrm{cm}^{-2}}
{N(\mathrm{H}_\mathrm{tot})}\right)
\left(\frac{0.08\,\mathrm{cm}^{-3}}{n(e)}\right)\, 710\, \mathrm{ppm}
\end{equation} 
where we have left out the less important terms for collisional de-excitation
and excitation by H$^0$.  This expression makes it clear that the C abundance
in the LIC is quite high unless $N($\ion{C}{2}$^*)$ is at the low end of its
range while $N(\mathrm{H}_\mathrm{tot})$ and $n(e)$ are the high ends of
theirs.  In this expression $n(e)$ should be the line of sight
averaged value, which is highest at the cloud edge and decreases inward on the
line of sight.  For our models we find that the expression leads to
about an 8\% overestimate for A$_\mathrm{C}$ when $n(e)$ is interpreted as the
electron density at the position of the heliosphere. If we take an upper limit
on $n(e) = 0.09$ \cc, consistent with our derivation above, and an upper limit
of $N(\mathrm{H}_\mathrm{tot}) = 8\times10^{17}$ \scm, and the
lower limit of $N($\ion{C}{2}$^*) = 1.1 \times 10^{12}$ cm$^{-2}$ we get 334
ppm as a lower limit on the C abundance. This lower limit is consistent with
the solar abundance from \citet{Grevesse+Sauval_1998}, but not with the more
recent results of AGS.  Using a value of
$N(\mathrm{H}_\mathrm{tot})$ as high as $8\times10^{17}$ implies a S abundance
of $A_\mathrm{S} \lesssim 13$ ppm to be consistent with the observed upper
limit for $N($\ion{S}{2}$)$.  Sulfur is expected to have little to no
depletion in warm clouds \citep[see e.g.,][]{Welty23:1999}, and a solar
abundance of 13.8 ppm is found by AGS.

\section{Discussion\label{sec:discussion}}

An implicit assumption in our analysis for the electron density is
photoionization equilibrium.  The timescale for \ion{Mg}{1}
ionization is very short, $< 10^3$ years, while the recombination timescale
for \ion{Mg}{2} is considerably longer $\sim 3\times10^5$ yr.  
The most likely scenario for the cloud to be out of photoionization
equilibrium is that it is recombining and cooling from a previously heated
(e.g., shocked) state \citep[see][]{Lyu+Bruhweiler_1996}. In that case the
electron density we infer would be an underestimate.  This would both lower
the inferred value of \ACAS\ and the C abundance.  
However, \citet{Jenkins_etal_2000b} argue, based on the relative column
densities of \ion{Ar}{1} and \ion{H}{1} in the LIC, that photoionization
dominates the ionization rather than the collisional processes that should
dominate in cooling and recombining gas. It remains to be seen whether or not
a non-equilibrium model could be found that would be consistent with the
\ion{Ar}{1} data as well as all the other column density data for the LIC and
the \emph{in situ} data on He$^{0}$. We intend to explore this possibility in
the future.

Observations of low column density, \NH $< 10^{19}$ \scm,
ISM towards other nearby stars, are consistent with our conclusion of
supersolar C abundances.  Using \NI+\NII\ as a proxy for the total H column
and the AGS Solar abundance for N for the lines of sight towards Capella,
WD~1254+223, WD~1314+293, and WD~1634-573 yields $A_\mathrm{C}\sim340$ ppm,
with a range $83-550$ \citep[see data in][]{Woodetal:2002capella,
Lehneretal:2003, Oliveira_etal_2005,Kruketal:2002}.
Three of these stars show $A_\mathrm{C}>333$ ppm.

\section{Conclusions}

We have shown that using updated atomic data and the \emph{in situ}
data on the temperature of the gas in the LIC, we can put interesting
constraints on the C abundance in the LIC.  Unless a number of observed or
inferred data values are at the far ends of their uncertainty ranges, it
appears that C is not only undepleted in the LIC, but has an abundance of
$A_\mathrm{C} \approx 400-800$ ppm, considerably larger than the solar
value.  In addition, the ratio of C to S abundance appears to be necessarily
larger than the solar ratio of $\sim 20$ with values in the range
of $40 - 50$ favored by the constraints on $n(e)$ in the LIC.  We speculate
that this supersolar abundance of C, if it truly exists, may be indicative of
a local enhancement in dust in the past.  In this picture the gas phase C
abundance was enhanced by the passage of an interstellar shock that destroyed
the carbon bearing grains more effectively than other, e.g.\ silicate, grains.
This could be related to the unusual dust size distribution in the LIC as
indicated by Ulysses observations of large grains penetrating into the
heliosphere \citep{Grun_etal_1994,Frisch_etal_1999}.

Further investigations into the C abundance in the LIC should be undertaken to
confirm our findings, especially studies of other lines of sight that show the
LIC velocity component in the absorption lines of C, S, Mg and other elements.
The LIC is the best observed low density interstellar cloud and thus can
provide us with valuable information on this phase of the interstellar medium.
The C abundance and its evolution in the warm diffuse ISM has many
implications for the nature of interstellar dust and metallicity in the Galaxy.

\acknowledgements
This research was supported by NASA grants NAG5-12571 and NAG5-13107.  
The authors thank Daniel Savin and Gary Ferland for very helpful discussions
and the anonymous referee who helped improve the clarity of the presentation.

%
% ---- Bibliography ----
%

\clearpage
\begin{deluxetable}{lcc}
\tablecolumns{3}
\tablewidth{0pt}
\tablecaption{Observational Constraints \label{tab:coldens}}
\tablehead{
\colhead{Observed} &  \colhead{Observed} &
\colhead{Reference} \\
\colhead{Quantity} &  \colhead{Value} & \colhead{} \\
}
\startdata
$N($\ion{C}{2}$)$ (cm$^{-2}$) & $1.4 - 2.1\times10^{14}$ & 1 \\
$N($\ion{C}{2}$^*)$ (cm$^{-2}$) & $1.3 \pm 0.2 \times10^{12}$ & 1 \\
$N($\ion{O}{1}$)$ (cm$^{-2}$) & $1.4^{+0.5}_{-0.2} \times10^{14}$ & 1\\
$N($\ion{Mg}{1}$)$ (cm$^{-2}$) & $7 \pm 2 \times10^{9}$ & 1 \\
$N($\ion{Mg}{2}$)$ (cm$^{-2}$) & $3.1 \pm 0.1 \times10^{12}$ & 1 \\
$N($\ion{S}{2}$)$ (cm$^{-2}$) & $8.6 \pm 2.1 \times10^{12}$ & 1 \\
$T$(K) & $6300 \pm 340$ & 2 \\
$n($\ion{He}{1}$)$ (cm$^{-3}$) & $0.015\pm 0.0015$ & 2 \\
\enddata
\tablerefs{(1) \citet{Gry+Jenkins_2001} (Values shown are for the LIC
component towards \epsC, component 1.), (2) \citet{Witte_2004}.}
\end{deluxetable}

\end{document}